\documentclass[aps,prl,reprint,groupedaddress]{revtex4-1}
\usepackage{graphicx}
\usepackage{color}
\usepackage{siunitx}

\renewcommand{\vec}[1]{{\mathbf #1}}

\begin{document}
	
\title{Dynamical defects in rotating magnetic skyrmion lattices}

\author{S. P\"ollath$^1$, J. Wild$^1$, L. Heinen$^2$, T.N.G. Meier$^1$, M. Kronseder$^1$,  L. Tutsch$^1$, A. Bauer$^3$, H. Berger$^4$, C. Pfleiderer$^3$, J. Zweck$^1$, A. Rosch$^2$,  C.H. Back$^1$}

\affiliation{$^1$ Institut f\"ur Experimentelle Physik, Universit\"at Regensburg, Universit\"atsstr. 31, D-93040 Regensburg, Germany,
                            $^2$ Institut f\"ur Theoretische Physik, Universit\"at zu K\"oln, D-50937 K\"oln, Germany, $^3$ Physik-Department, Technische Universit\"at M\"unchen, D-85748 Garching, Germany, $^4$ Crystal Growth Facility, \'Ecole Polytechnique F\'ed\'erale de Lausanne, CH-1015 Lausanne, Switzerland}
	
\date{\today}

	\begin{abstract}{The chiral magnet Cu$_{2}$OSeO$_{3}$  hosts a skyrmion lattice, that may be equivalently described as a superposition of plane waves or lattice of particle-like topological objects. A thermal gradient may break up the skyrmion lattice and induce rotating domains raising the question which of these scenarios better describes the violent dynamics at the domain boundaries. Here we show that in an inhomogeneous temperature gradient caused by illumination in a Lorentz Transmission Electron Microscope different parts of the skyrmion lattice can be set into motion with different angular velocities. Tracking the time dependence we show that the constant rearrangement of domain walls is governed by dynamic 5-7 defects arranging into lines. An analysis of the associated  defect density is described by Frank's equation and agrees well with classical 2D-Monte Carlo simulations. Fluctuations of boundaries show surge-like rearrangement of skyrmion clusters driven by defect rearrangement consistent with simulations treating skyrmions as point particles. Our findings underline the particle character of the skyrmion.}
\end{abstract}

\maketitle

In the last decade a non collinear topological spin texture, the skyrmion, has attracted great attention representing a new type of topological soliton in magnetic materials. Skyrmion lattices are periodic arrangements of a kind of magnetic whirls that may be found in a great variety of chiral magnets~\cite{1989:Bogdanov,1994:Bogdanov,2009:Muhlbauer:Science,2010:Munzer:PhysRevB,2010:Pfleiderer:JPhysCondensMatter,2010:Yu:Nature,2011:Yu:NatureMater,2012:Seki:Science,2013:Milde:Science,2015:Tokunaga:NatCommun,2013:Nagaosa:NatureNano,2016:Bauer:Book}, as well as thin magnetic (multi-) layers  \cite{2013:Romming:Science,2016:Woo:NatMater,2016:Boulle:NatNano,2016:MoreauLuchaire:NatNano}. The topology of skyrmions is encoded in a quantized winding number of the spin orientation.
Emergent magnetic and electric fields describe the efficient coupling of the topological spin texture to electrons and magnons
~\cite{2009:Neubauer:PhysRevLett, 2010:Jonietz:Science, 2012:Schulz:NaturePhys, 2012:Yu:NatCommun}.

Skyrmion lattices may be described by two distinct approaches. In a wave-like picture as suggested e.g. by  Small-Angle-Neutron Scattering (SANS) measurements, the skyrmionic crystal may be accounted for by a superposition of three spin helices with their propagation vector rotated by $120^\circ$ with respect to each other.
From this point of view,
the individual (solitonic) character of single skyrmions vanishes in the collective. 
Note that in most materials in the skyrmion lattice phase,  higher order scattering is basically absent; for MnSi it is of the order of 10$^{-4}$ suggesting a rather smooth spin texture \cite{2011:Adams:PhysRevLett}.
In contrast, in a particle-like picture skyrmions are viewed as individual solitonic particles.
Indeed, individual skyrmions have been observed early on \cite{2010:Yu:Nature,2013:Romming:Science} but the non-linear character as well as the degree of the particle-character in the skyrmion phase remained unresolved. In fact, recent studies reveal strong deformation of the precise shape of skyrmions under large strain \cite{Shibata2015}.

The validity of either approach may be tested critically in studies of imperfect skyrmion lattices, 
alluding to similarities with well-known atomic lattices which also allows to verify particle conservation.
In fact, the existence of defects and domains in skyrmion lattices
has been reported \cite{boundaryJapan,boundaryPNAS,2016:Zhang:NanoLett,boundaryDPC}, however, no comprehensive study focused on their dynamical properties has been performed, yet. 
Here, we use time resolved non-stroboscopic cryo-Lorentz transmission electron microscopy (L-TEM) to record domain boundaries and defects along with their fluctuations in the skyrmion lattice phase of mechanically thinned Cu$_{2}$OSeO$_{3}$ single crystals and find strong indications for the  particle-like character of skyrmions in this
material. We analyze the motion of single defects and of grain boundaries in the skyrmion lattice
and perform a large statistical analysis of the defect density for a rich selection of misorientation angles $\Delta\theta$ between the skyrmion domains.

	\begin{figure*}[tb]
		\includegraphics[width = 1.0\textwidth]{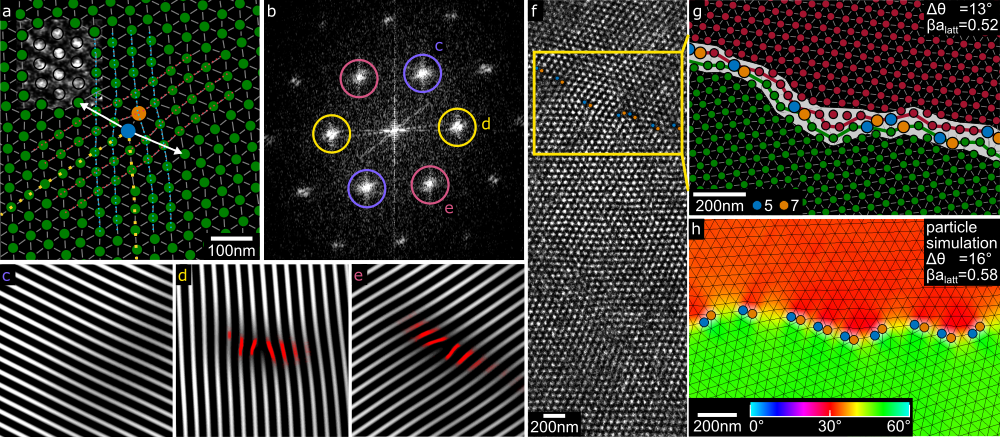}	
		\caption{\label{figure_sigBoundaries} (a) 5(blue)-7(orange)-defect with two disclination lines (yellow dotted lines). White arrows indicate preferred fluctuation directions. The grey inset shows L-TEM data. (b) Fourier transformation of (a). Colored circles represent masks to obtain images (c)-(e) after an inverse FT. (c)-(e) Inverse FTs of individual pairs of spots seen in (b). The red color encoding corresponds to the lattice deformation. (f) Background subtracted L-TEM data. The yellow box is enlarged in (g).
			(g) Grain boundary in the skyrmionic lattice. The panel displays processed data where skyrmions with coordination number 5 (7) are marked with blue (orange) dots. The two grains of  the skyrmion lattice are marked with red and green dots. The red and green lines are estimations of the grain boundary detected by an algorithm. (h) Grain boundary obtained by particle simulations. The color encodes local orientations of the skyrmion lattice  \cite{supplement}.}
	\end{figure*}

As our main result we find skyrmion domains with high mobility  as well as simultaneous surge like rearrangement of many skyrmions triggered by magnon currents due to the temperature gradient introduced by the electron beam. 
The fluctuating domain boundaries may be described by a combination of 5-7 defects while the defect density increases with the misorientation angle between the grains. The observed domain boundaries are similar to 2D dense packed lattices of colloidal crystals, bubble rafts or even the nanonipple structure on the facet eyes of the morning cloak butterfly \cite{colloidalDefect, BraggNye1947, boundaryButterfly} which is rather spectacular since the latter systems consist of solid particles while the skyrmion is a continuous spin texture.

Our findings are supported by 2D Monte Carlo simulations of the spin texture and by particle-based simulations of the dynamics.
 They allow insight into the nature of this peculiar lattice of topological spin objects and emphasize its
 particle-like character. Particle-like properties are important and desired since future spintronic applications aim at controlling single skyrmions \cite{2013:Fert:NatNano}. 

For our study we have selected the multiferroic insulator Cu$_{2}$OSeO$_{3}$, which is a prominent example for a bulk material hosting a skyrmionic crystal~\cite{2012:Seki:PhysRevB2} as observed in reciprocal space using SANS~\cite{2012:Adams:PhysRevLett, 2012:Seki:PhysRevB, 2014:White:PhysRevLett} and X-ray magnetic scattering~\cite{2014:Langner:PhysRevLett, 2016:Zhang:PhysRevB, 2016:Zhang:NanoLett} and in real space via L-TEM~\cite{2012:Seki:Science} and MFM~\cite{2016:Milde:NanoLett}.
Defects of the skyrmion phase of Cu$_{2}$OSeO$_{3}$ were reported first by Rajeswari et $\it{al}$. \cite{boundaryPNAS} and Zhang et \textit{al.}\ who studied a multi-domain state by resonant X-ray scattering \cite{2016:Zhang:NanoLett}, however, without further details.

The Cu$_{2}$OSeO$_{3}$ single crystal was grown by the chemical vapor transport method and it was thinned in the $[110]$ direction first mechanically and then by argon-ion milling technique down to a thickness of $\approx 100\,\text{nm}$. 
A few nanometers of carbon were evaporated onto the thinned sample to minimize charging effects.
L-TEM images were taken with a Tecnai F30 TEM in Lorentz mode where the magnetic field normal to the sample surface is tuned by the objective lens current. A defocused image with the skyrmions is projected on a phosphorescence screen, recorded through a lead glass window by a digital camera with a maximum of 90 frames per second. We controlled the temperature of the sample with a Gatan liquid helium sample holder. All data shown here was recorded using a constant temperature and externally applied magnetic field of $18\,\text{K}$ and $48\,\text{mT}$. The phase diagram was entered by zero field cooling to $18\,\text{K}$ and subsequently applying a magnetic field until the skyrmion lattice phase is present in the whole field of view.

In a perfect skyrmion lattice phase each skyrmion is surrounded by 6 neighbors, which does not hold when differently oriented skyrmion lattice domains are present.
In our experiments and simulations we observe that the static and dynamic properties of domain  boundaries are best described by 5-7 defects, i.e., two skyrmions which have 
five or seven nearest neighbors, respectively. The coordination number of the defects depends on the geometry of the magnetic lattice as can be seen from a comparison to recent investigations by Hagemeister et al. concerning defects in a skyrmion square lattice \cite{HagemeisterIaiaVedmedenkoEtAl2016}. A single 5-7 defect is shown in  Figure \ref{figure_sigBoundaries}a.
The grey scale image in the upper left hand corner displays the raw L-TEM contrast after removal of image distortion, for details concerning image processing see Supplementary Material \cite{supplement}.

The Delaunay triangulation method is used to identify nearest neighbors.
A closer look reveals the true nature of this defect: two disclination lines (yellow dotted lines)  join in the lattice node with five neighbors. Figures \ref{figure_sigBoundaries}c-e  show inverse Fourier-Transformations (FTs) of individual pairs of spots of the skyrmion lattice's typical six spot FT pattern displayed in Figure \ref{figure_sigBoundaries}b. The disclination lines can be identified directly as extra lines in Figures \ref{figure_sigBoundaries}d,e ending at the defect position. Deformations (red color) occur predominantly perpendicular to the disclination lines.
Figure \ref{figure_sigBoundaries}c shows no significant deformation which means that the lattice along this third spin-crystallographic direction is not distorted. Note a slight lattice bending of around $3^\circ$ that is compensated by the 5-7 defect.  

In the following we explore the possible particle-like character of skyrmions by analyzing the boundaries of domains in motion. When skyrmionic domains move, their constituents must rearrange which allows testing particle conservation.
Early experiments \cite{2010:Jonietz:Science} showed that the combination of an electric current and thermal gradients can induce a rotation of the skyrmion lattice by finite angles \cite{EverschorPRB2011,EverschorPRB2012}. Continuous, ratchet-like motion induced by thermal gradients from the electron beam of a TEM was observed
by Mochizuki et $\it{al.}$ \cite{TokuraRatchetMotion}.
The temperature gradient induced by the electron beam is not homogeneous  leading to different effective torques at different positions of the sample and consequently causing parts of the skyrmion crystal to rotate at different angular velocities. Also homogeneous heat currents in combination with variations of the sample thickness can cause rotational torques \cite{EverschorPRB2011,EverschorPRB2012}.
 This eventually leads to domain formation and to heavy fluctuations of the domain boundaries, enabling us to record many different boundary configurations and tilt angle configurations. 

	\begin{figure}[tb]
		\includegraphics[width = \linewidth]{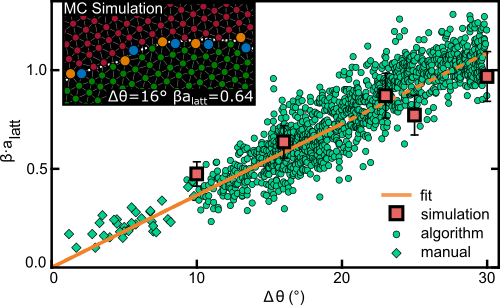}%
		\caption{\label{figure_histo} Defect densities along the grain boundaries relative to the misorientation angle evaluated manually and by an algorithm. Also shown are values from Monte Carlo simulations and a fit formula for symmetric low angle grain boundaries. The inset shows an exemplary Monte Carlo simulation for a tilt angle of $\Delta\theta=15^\circ$.  }
	\end{figure}

Skyrmion crystallites with different lattice orientations require that defects form a grain boundary and the 5-7 defect is the disclination of choice for the skyrmionic lattice.
For low tilt-angles between skyrmion crystallites single 5-7 defects line up to compose the boundary. Large angles ($>20^\circ$) produce a boundary that mainly consists of strictly alternating nodes with coordination numbers 5 and 7. 
	
Figure \ref{figure_sigBoundaries}f shows
a grain boundary for a tilt angle of $13^\circ$, other examples are shown in the Supplementary Material  \cite{supplement}.
The panel shows a processed image highlighting the two domains and their boundary region. The positions of individual skyrmions
are marked as red and green dots, skyrmions with coordination number 5 (7) blue (orange).
The grey rectangular area underneath parts of the processed data display the respective raw L-TEM image.

Importantly, we observe no preferred tilt angle in a statistical analysis of fluctuating domain walls that occupy many different configurations, see Fig.~\ref{figure_histo}. Apparently the energy scale of a grain boundary as a function of misorientation angle is much smaller than the energy scales dominating the domain and defect motion \cite{GottsteinBook2010}. 
Furthermore, the skyrmion lattice is not pinned to the atomic crystal and freely floats on the sample. 
	
The defect density $\beta$ along the border for the detected tilt-angles is shown in figure \ref{figure_histo}. In order to improve statistics for the analysis of defect densities along domain boundaries, an algorithm was developed that detects a potential lattice tilt by searching peaks in the histogram of nearest neighbor angles with respect to a fixed axis. Using this information it is possible to automatically detect the grain boundary and roughly measure its length, while manual measurements are necessary for low angles ($<10^\circ$). 

In solid crystals  Frank's equation accounts for the spacing $d$ of individual defects in relation to the misorientation angle of grains and holds for symmetric small-angle grain boundaries \cite{FranksFormula}.
 Provided that the length of the Burgers vector of the 5-7 defect is one skyrmion lattice constant $a_\text{latt}$, it can be rewritten as $\beta = \frac{2}{d} = \frac{4\cdot\sin\left(\Delta\theta/2\right)}{a_\text{latt}}$. A fit of the data to this equation in the range up to $20^\circ$ converges with $\frac{a_\text{latt}}{a_\text{latt,measured}} = 0.95 \pm 0.08 $ and is in good agreement with the data. The dashed line is the extension of the fit to larger angles, where it still matches the data well.
	
To describe our findings taking into account the underlying spin texture, classical 2D Monte-Carlo Simulations were performed following  \cite{2010:Yu:Nature}. The results of these simulations are shown in Figure~\ref{figure_histo}.
To maintain a boundary spins were fixed at the simulation grid’s edges forming starting points for two lattices that are tilted with respect to each other. Simulated annealing results in a grain boundary consisting of 5-7 defects in the grid’s center similar to our experimental observations. An example for a misorientation angle of 15$^\circ$ can be seen in the inset of Fig.~\ref{figure_histo}. The line-defect-densities of these simulations were manually evaluated and also included in Fig.~\ref{figure_histo}. Additional information can be found in the Supplementary Material \cite{supplement}.

Observations of the boundaries' dynamics show surge like rearrangement of Skyrmion clusters changing their affiliation to an individual domain between frames. Row shifts and small rotating groups of skyrmions can be seen as mechanisms of affiliation change. The skyrmion number during these fluctuations is conserved  with an accuracy of $\pm$ 1 in the analyzable image sections, supporting the picture of single topological particles that rearrange. This strongly hints to topological charge conservation during the rearrangement processes.

	\begin{figure}[b]
		\includegraphics[width = \linewidth]{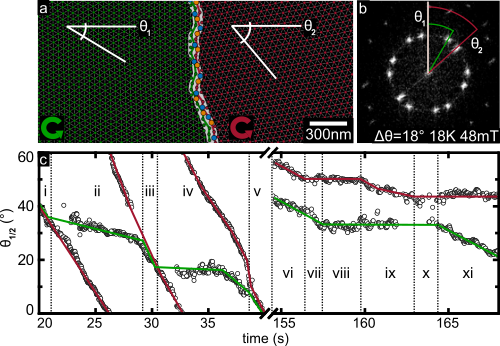}%
		\caption{\label{figure_evol} 
			Temporal evolution of the angles $\theta_1$ and $\theta_2$ of the grains relative to a fixed axis. (a) displays the two grains and the measured angles. The circular arrows represent the rotational direction. (b) is a Fourier transformation of the data, where $\theta_1$ and $\theta_2$ are also present. The lower panel (c) shows two sequences in the temporal evolution of $\theta_1$ and $\theta_2$. In the first sequence one grain rotates faster than the other going four times through the full range of misorientation. In the second sequence variations of individual rotational speeds lead to increase and decrease of misorientation.}
	\end{figure}

The temporal evolution of the misorientation angle in Fig. \ref{figure_evol} demonstrates the influence of the electron beam on two domains of the lattice.
To visualize their dynamics we record a 170~s long movie using a frame rate of 30 fps. The sequence shown in the Supplementary Movie \cite{supplement} ranges from about 20~s to 30~s. The left and right parts of Fig. \ref{figure_evol}c show the two angles of the rotating domains from 20~s to 40~s and from 155~s to 170~s, respectively. The angles $\theta_1$ and $\theta_2$ can consistently be obtained either from the real-space images or from its Fourier transform, see figures \ref{figure_evol}a,b.
The evaluation of the first sequence reveals a fast rotation of one lattice (red)  while the other (green)  rotates much slower. It captures four full passes through all possible relative domain angles. For vanishing relative orientations (timespans i, iii,v) the domain wall dissolves and a unified lattice is obtained, i.e. there is no grain boundary or only single defects visible in the field of view.

The second sequence shows a different process. At first both grains rotate with almost the same velocities keeping the misorientation angle constant (vi). Next, one grain stops rotating  leading to an increase of the net misorientation angle (vii). Before the maximum misorientation of $\Delta\theta=30^\circ$ is reached the other grain also stops
(viii). During time span (ix) the inverse process of (vii) occurs, effectively reducing the misorientation angle. Now both grains stop (x) before one grain starts to move again (xi). 

 Note that the rotation direction during these processes remains mainly fixed in counter clockwise direction in accordance with the findings of reference~\cite{TokuraRatchetMotion}. However, slight deviations of this strict sense of rotation are observed for instance in region (vii), which we verified by raw data. This rotation in opposite direction can be interpreted as gearwheel rotation of both lattices. The relative angular speed varies from $0$ to $15^\circ/\text{s}$.
The different behavior of the two sequences indicates that the effective rotational torques are substantially larger during the first sequence compared to the second sequence. As the temperature gradients are expected to be approximately constant in time, this arises very likely from changes in the size of the domains and therefore in the total effective forces at the domain boundary.
Indeed, indicated by a tilt of the orientation of the grain boundary, the relative domain sizes changed in between the two sequences.
\begin{figure}
	\centering
	\includegraphics[width=\linewidth]{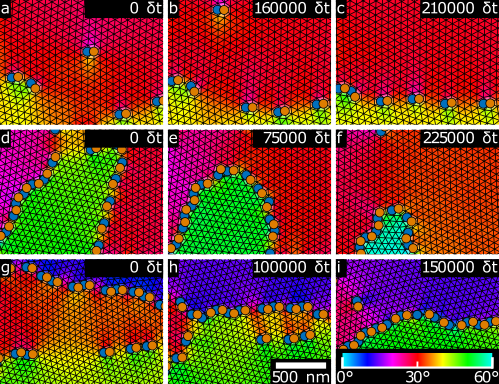}
	\caption[Particle based simulations]{Snapshots of dynamic simulation of skyrmion domains rotating due to thermal currents. As in Fig.~\ref{figure_sigBoundaries}(h) defects with coordination number 5 or 7 are marked as thick circles. Panels (a)-(i) show some of the processes important for domain formation. (a)-(c) 5-7 defects move to reduce strain and align into a domain wall. (d)-(f) Rearrangement of domain walls accompanied by the annihiliation of defects. (g)-(i) Two domain walls merge.  }
	\label{figure_sim}
\end{figure}

The topological protection  of skyrmions,  the associated approximate conservation of skyrmion number observed in our experiment and the dominance of 5-7 defects suggest that skyrmions should be viewed as discrete particles even in lattice textures. To corroborate this picture we have performed simulations of large rotating skyrmion lattices and track the motion of $\sim 35.000$ skyrmions treated as individual particles \cite{2013:Lin:PRB} arranged on a disk, see Supplementary Material \cite{supplement}. 
Even for this simplified approach it is not possible to simulate directly the experimentally relevant time scales of tens of seconds which is a factor $10^{10}$ larger than the microscopic time scale for typical excitations of the skyrmion lattice \cite{2015:Schwarze:NatMater}. Remarkably, we are, however, able to reproduce qualitatively the experimentally observed dynamics using current densities much larger than in the experiment.

Under the shear stress arising from the current pattern, domains characterized by different rotational angle form. Most importantly, the simulations show the dynamical formation of domain walls consisting of 5-7 defects similar to our experimental observations, see Fig. \ref{figure_sigBoundaries}g and \ref{figure_sim}. 
The domain dynamics is a highly dynamic process characterized by continuous creation and destruction of domains. Within our simulation the rapid motion of single 5-7 defects is a dominating mechanism for the change of defect density, and thus the relative rotation of domains. We observe strongly fluctuating domain sizes, domain mergers and surge-like rearrangements of domains in our simulations, see Fig.~\ref{figure_sim}. The Supplementary Movie gives an overview of the dynamical, highly non-linear processes which accompany the rotation of the domains\cite{supplement}.

In summary, we observe extraordinary dynamics of rotating skyrmion domains while exposed to a temperature induced torque. The dynamics are largely dominated by the motion and rearrangement of 5-7 defects, whose density at domain walls has to adjust to the continuously changing misorientation angle of neighboring domains. The domain boundary shows heavy fluctuations but ultimately allows for a continuous net rotation of individual domains driven by the applied torques. 
Our findings support the skyrmion's particle-like character in the material Cu$_{2}$OSeO$_{3}$.
The particle-like character of skyrmions is evidenced by the existence of 5-7 defects, the correspondence of the defect density along the grain boundary with classical particle theory and the particle-like rearrangement mechanisms during grain boundary fluctuations. 
Furthermore external manipulation of individual skyrmion grains is possible by applying a temperature gradient.


%


\clearpage

\onecolumngrid
\appendix

	\begin{center}\large{Dynamical defects in rotating magnetic skyrmion lattice}\\ \Large{\textbf{Supplementary Material}}\end{center}

	\subsection{Supplement 1: Image processing}
	
	\begin{figure}[!b]
		\centering
		\includegraphics[width=0.8\linewidth]{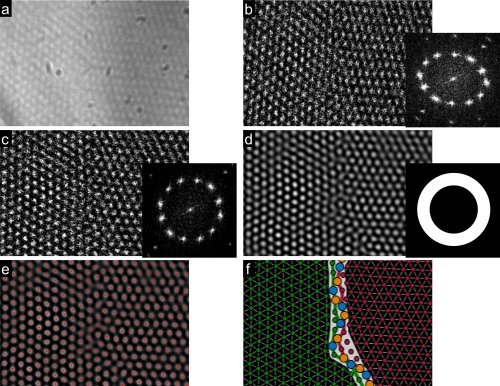}
		\caption[Image processing]{
			Image processing (a) Raw Lorentz-TEM image. (b) Background subtraction from video averaging. The inset shows the distorted pattern of the Fourier transformation. (c) Distortion compensated image. The Fourier transformed pattern in the inset is now circular. (d) Bandpass filtering increases the signal-to-noise ratio further. (e) Skyrmion positions detected from (d) and marked with red dots. (f) Detection of grains, grain boundaries and defects.}
		\label{fig_01ImageProcessing}
	\end{figure}
	
	To detect skyrmion positions advanced image processing of the data is necessary. First the raw sequence of images (Fig. \ref{fig_01ImageProcessing}a) is referenced to compensate image drifts. Subsequently background needs to be removed. For this the average intensity of a video sequence of approximately three seconds is taken. 
	Due to its fast rotation the magnetic contrast of the skyrmion lattice averages to a constant value while the image background remains unchanged. After subtraction (Fig. \ref{fig_01ImageProcessing}b), distortion of the image is compensated. This distortion originates from the measurement technique of recording a phosphorescent screen which is located inside the transmission electron microscope (TEM) so that it is canted with respect to the incident electron beam. The distortion can be seen in the Fourier transformation (inset Fig. \ref{fig_01ImageProcessing}b) as a canted ellipse of the typical six spot pattern. To get rid of this distortion the image is rotated such that the main axes of the ellipse lie parallel to the rectangular pixel grid and then scaled to best possibly reduce the ellipticity of the regularly round spot pattern (cross-checked using undistorted images obtained via the built-in CCD camera). The undistorted image and the round FFT spot pattern can be seen in Fig. \ref{fig_01ImageProcessing}c. To enhance the signal-to-noise ratio further the images are bandpass filtered (Fig. \ref{fig_01ImageProcessing}d, filter in the inset). Next the skyrmion positions are detected using a local maximum finding algorithm (Fig. \ref{fig_01ImageProcessing}e). Finally Delaunay triangulation can be used to detect nearest neighbors and from that grains, grain boundaries, misorientation angles and defects in the skyrmion lattice (Fig. \ref{fig_01ImageProcessing}f).

	\subsection{Supplement 2: Examples for CSL grain boundary configurations}
	
	\begin{figure}
		\centering
		\includegraphics[width=1.0\linewidth]{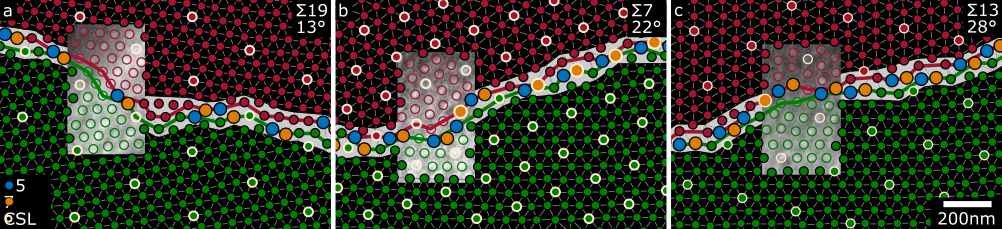}
		\caption{(a) $\Sigma$-19, (b), $\Sigma$-7 and (c) $\Sigma$-13 grain boundaries in the skyrmion lattice. 	
			The panels show processed data where skyrmions with coordination number 5 are marked with blue dots and with coordination number 7 with orange dots. The grey rectangles show the raw L-TEM images. 	
			The two grains of  the skyrmion lattice are marked with red and green dots. The red and green lines are the respective boundary estimations detected by an algorithm. White rings mark the corresponding coincidence site superlattice (CSL).}
		\label{fig:02-GrainBoundaries}
	\end{figure}
	
	As mentioned in the main text, many different grain boundary configurations were recorded. Fig. \ref{fig:02-GrainBoundaries} shows a selection of grain boundaries for tilt angles of $13^\circ$ , $22◦^\circ$ and $28◦^\circ$. The panels show processed images highlighting the two crystallites and their boundary region. Here, the positions of the individual skyrmions are encoded as red and green dots in the two crystallites. A region of interest for the boundary is shown in grey and skyrmions with coordination number 5 (7) in blue (orange). Red and green lines are the estimated boundaries. Grey rectangular areas underneath parts of the processed data show the raw distortion compensated L-TEM images. As can be seen in the images, the defect density rises as a function of tilt-angle. Furthermore, the boundary is not straight but shows meander-like progression. For the three examples shown in Fig. \ref{fig:02-GrainBoundaries} we also show the corresponding CSL-lattices (white rings around the positions of the skyrmions) revealing the $\Sigma$-values of these special boundaries as the ratio of the superlattice shared by both grains to the common lattice unit cell volume. We find that the shown grain boundaries can be described by (a) $\Sigma$-19, (b) $\Sigma$-7 and (c) $\Sigma$-13.
	
	\begin{figure}
		\centering
		\includegraphics[width=1.0\linewidth]{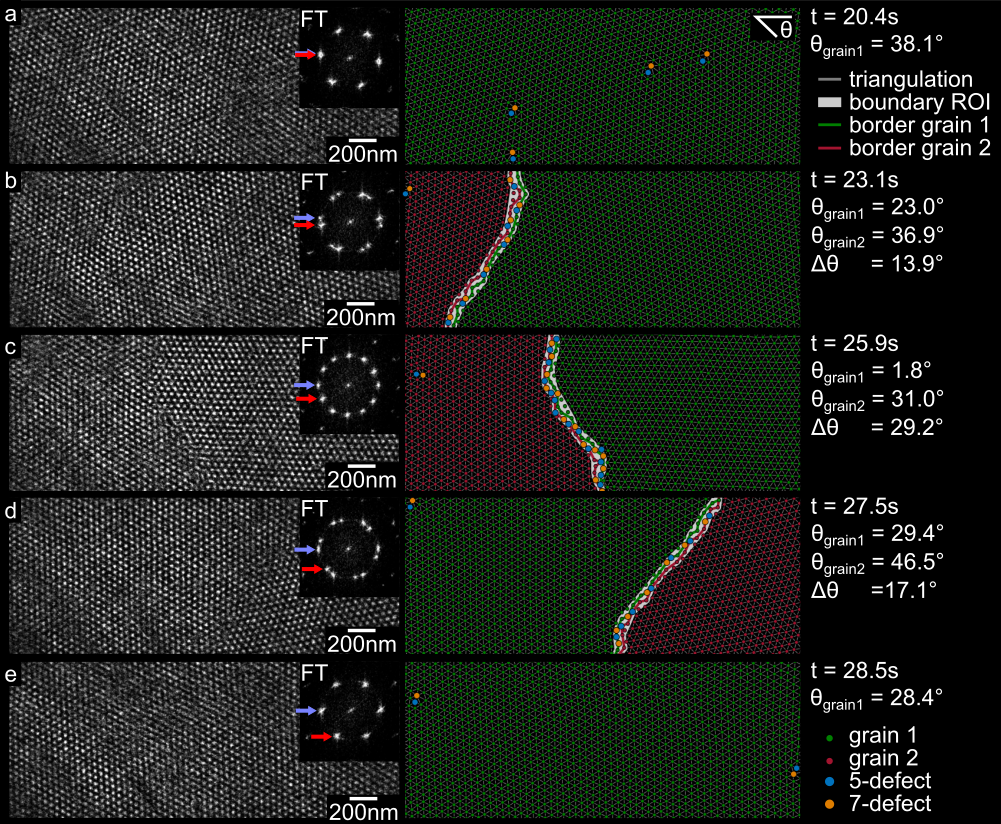}
		\caption[Creation and annihilation of a grain boundary]{Still images from the supplementary movie. The creation and annihilation of a grain boundary is shown. Each panel shows background subtracted data with the respective Fourier transformation as an inset and the processed data. (a) A nearly perfect hexagonal skyrmion lattice with four 5-7 defects is present. (b) The grains separate and a grain boundary forms. (c) The misorientation angle reaches its maximum value of $\approx 30^\circ$. (d) Misorientation angle and defect density along the grain boundary decrease again. (e) The grain boundary dissolves and a nearly perfect hexagonal lattice is obtained.}
		\label{fig_movieSnapshots}
	\end{figure}
	
	\subsection{Supplement 3: Capturing the dynamics of a grain boundary under grain rotation}
	
	A movie is presented to demonstrate the magnon-current driven rotation of the skyrmion grains. In the video the two grains rotate with different angular velocities leading to an increase of misorientation angle from $0^\circ$ up to $30^\circ$ and a subsequent decrease to $0^\circ$. Fig. \ref{fig_movieSnapshots} shows five still images of the video. Each panel contains background subtracted data on the left with the respective Fourier transformation in an inset and the processed data on the right. Note that in Fig. \ref{fig_movieSnapshots} as well as the movie, 5-7 defects next to the grain boundary were marked manually while omitting possible defects that were detected from the algorithm but are located inside blurred regions of the raw data. Further given are time and grain orientation angles. The sequence starts with a nearly perfect hexagonal skyrmion lattice with few 5-7 defects present (Fig. \ref{fig_movieSnapshots}a). A clear grain boundary forms and the misorientation angle increases to $\approx 30^\circ$ (Fig. \ref{fig_movieSnapshots}b,c). Note the separating spots in the respective FT images. Finally the misorientation angle decreases again (Fig. \ref{fig_movieSnapshots}d) until the misorientation approaches $0^\circ$ and there is no grain boundary visible inside the field of view (Fig. \ref{fig_movieSnapshots}e).

	\subsection{Supplement 4: Monte Carlo Simulations}
	
	\begin{figure}
		\centering
		\includegraphics[width=0.8\linewidth]{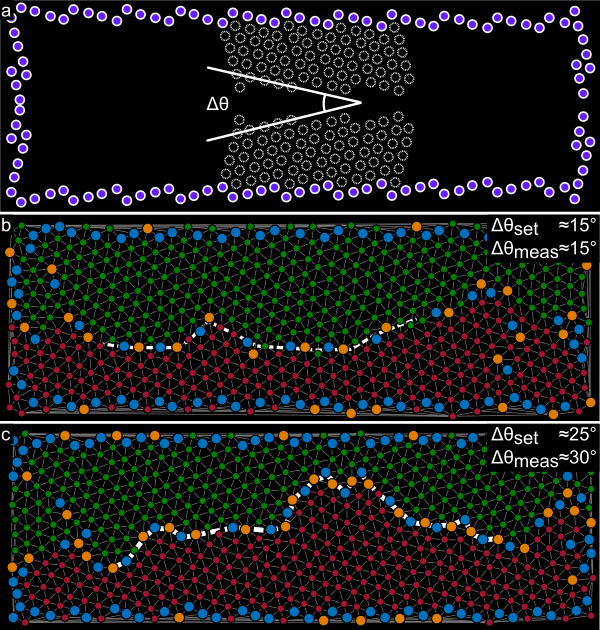}
		\caption[Monte Carlo Simulations]{Classical 2D Monte Carlo simulations for skyrmion grain boundaries. (a) Mask of fixed spins: Purple dots show fixed spins antiparallel to the external magnetic field. White dashed rings are the expected extension of the skyrmion lattice after relaxation. In this way the misorientation $\Delta\theta$ can be set. (b,c) Skyrmion positions of the relaxed grid after simulated annealing. Grain affiliations (green, red) and grain boundaries (white) are evaluated manually. Skyrmions with 5 neighbors are marked in blue and ones with 7 neighbors in orange as in the main text. Defects at the borders appear naturally and can be neglected. In the center of the grid a grain boundary forms which is lost near the short grid edges. The preset misorientation $\Delta\theta_\text{set}$ is not perfectly followed by the simulation for larger misorientation angles and results in  $\Delta\theta_\text{meas}$ (c, see main text). }
		\label{fig_02MonteCarlo}
	\end{figure}

	For the Classical Monte Carlo simulations we follow the work of Tokura et. $\it{al}$ \cite{2010:Yu:Nature}. Simulated annealing was used to relax Heisenberg spins on a 2d-plane ($\hat{x}$-$\hat{y}$) using a single flip Metropolis algorithm with the lattice Hamiltonian \cite{2010:Yu:Nature}
	\begin{equation}
	H_{2D} =  \sum_{\vec{r}} \left[- J  \vec{S}_{\vec{r}} \cdot \left( \vec{S}_{\vec{r}+a\hat{x}} + \vec{S}_{\vec{r}+a\hat{y}} \right) -K  \left( \vec{S}_{\vec{r}} \times \vec{S}_{\vec{r}+a\hat{x}} \cdot \hat{x} + \vec{S}_{\vec{r}} \times \vec{S}_{\vec{r}+a\hat{y}} \cdot \hat{y} \right) - \vec{H} \cdot \vec{S}_{\vec{r}}    \right] \text{ .}
	\end{equation}
	Here ferromagnetic exchange $J$, Dzyaloshinskii-Moriya interaction $K$ and Zeeman coupling to an external magnetic field $\vec{H}$ are considered. Discretization takes place on a 2D square lattice with lattice sites $\vec{r}$ lattice constant $a$ along the base vectors $\hat{x}$ and $\hat{y}$. The spins $\vec{S}$ were restricted to a unit sphere such that $\left|\vec{S}\right|=1$.
	
	First for fixed parameters $J$ and $K$ a temperature-field phase diagram is obtained by gradually lowering $T$ with a relaxation time of $\SI{e5}{}$ lattice sweeps between temperature steps on a 30x30 lattice with periodic boundary conditions until $T/J=0.1$. This supplies the information of suitable field values for a skyrmion lattice ground state. The ratio of $J$ and $K$ was chosen such that the spiral propagation length is 10 lattice sites.
	
	The simulation grid is then extended to a 360x120 lattice with open boundary conditions. Single spins near the long edges are fixed antiparallel to the applied external field in $\hat{z}$-direction. The fixed spins are chosen such that they make up a hexagonal lattice with a lattice constant measured from the skyrmion lattice of previous simulations. This artificial lattice is however tilted for opposing long grid edges in order to simulate any desired misorientation angle $\Delta\theta_\text{set}$. An example can be seen in Fig. \ref{fig_02MonteCarlo}a. The purple dots show the actual position of single fixed spins while the white dashed rings are an extension to show the lattice that is expected to form.
	
	As before simulated annealing is used to determine the ground state while applying an external magnetic field at which skyrmions are expected to occur from the phase diagram obtained before. Indeed, skyrmions are established around almost all of the fixed spins. Fig. \ref{fig_02MonteCarlo}b,c shows two examples of relaxed ground states for set misorientation $\Delta\theta_\text{set}=15^\circ$ and $\Delta\theta_\text{set}=25^\circ$. In the center of the grid where no spins are fixed the two grains in fact form a grain boundary comprising the defects described in the main text. Towards the short grid edges the clear boundary is lost which is due to the proximity to the border and formation of small additional grains. For larger misorientation angles the simulation does not exactly follow the configured tilt (e.g. $\Delta\theta_\text{set}=25^\circ$ results in $\Delta\theta_\text{meas}=30^\circ$ in Fig. \ref{fig_02MonteCarlo}c) which may be due to the fact that lateral movement of the grains is permitted by the fixed lattice positions. For these cases the measured misorientation was used for the defect density evaluation of the main text.

	\subsection{Supplement 5: Particle-based simulation and rotational dynamics}

	To model the formation and dynamics of rotating skyrmion lattices we treat each skyrmion as an individual classical particle with coordinates $\vec R_i$. Compared to direct micromagnetic simulations of spins this allows to treat much larger systems and longer time scales, speeding up simulations by more than 2 orders of magnitude. Furthermore, it allows us to find out whether a particle-based model correctly reproduces the experimental finding. In our simulations we model $\sim 35,000$ particles arranged on a disk. Even for this simplified approach it is not possible to simulate directly the experimentally relevant time scales of tens of seconds which is a factor $10^{10}$ larger than the microscopic time scale for typical excitations of the skyrmion lattice \cite{2015:Schwarze:NatMater}. Remarkably, we are, however, able to reproduce qualitatively the experimentally observed dynamics using current densities much larger than in the experiment, see below. The equations of motion are given by
	\begin{eqnarray}
	\vec{G} \times  \left( \dot \vec{ R}_i-\vec{v}(\vec{R}_i) \right)+\alpha \mathcal D 
	\dot \vec{ R}_i-\beta \mathcal D\vec{v}(\vec{R}_i)=\sum_j \vec F_{ij}
	\end{eqnarray}
	and solved by a straightforward Runge-Kutta approach.
	The dynamics of the skyrmions is dominated by the strong gyrocoupling which is fixed by the topology of the skyrmions, $\vec G\approx4 \pi m$, where $m$ is the two-dimensional spin density.
	The damping constant $\alpha \mathcal D$ is much weaker, using $\mathcal D/G=1.2$ we set $\alpha =0.05$ in our simulations. The terms proportional to $\vec{v}(\vec r)$ models the coupling of the skyrmion to heat currents \cite{Schroeter} and describes the velocity of spin-currents induced by the heat currents. We assume a Gaussian beam profile $\sim e^{-(r/r_0)^2/2}$ and use the two-dimensional continuity equation to determine the corresponding heat current profile. Using that $\vec v$ is proportional to the heat current we obtain $\vec v(\vec r)=-\frac{\vec r}{r^2}  v_0 (1-e^{-(r/r_0)^2/2})$. Note that $\vec v$ is oriented {\em towards} the center reflecting that a single skyrmion tends to move {\em towards} a heat source \cite{2013:Kong:PRL, SchuetteGarst2014, 2013:Nagaosa:NatureNano,Schroeter}. Theoretical calculations \cite{Schroeter} suggest that $\beta$ is negative, much larger than $\alpha$, but smaller than $1$. We set $\beta=-0.3$ for our calculation but have checked that  qualitatively similar results are obtained for $\beta=0$. 
	
\begin{figure}
		\centering
		\includegraphics[width=0.7\linewidth]{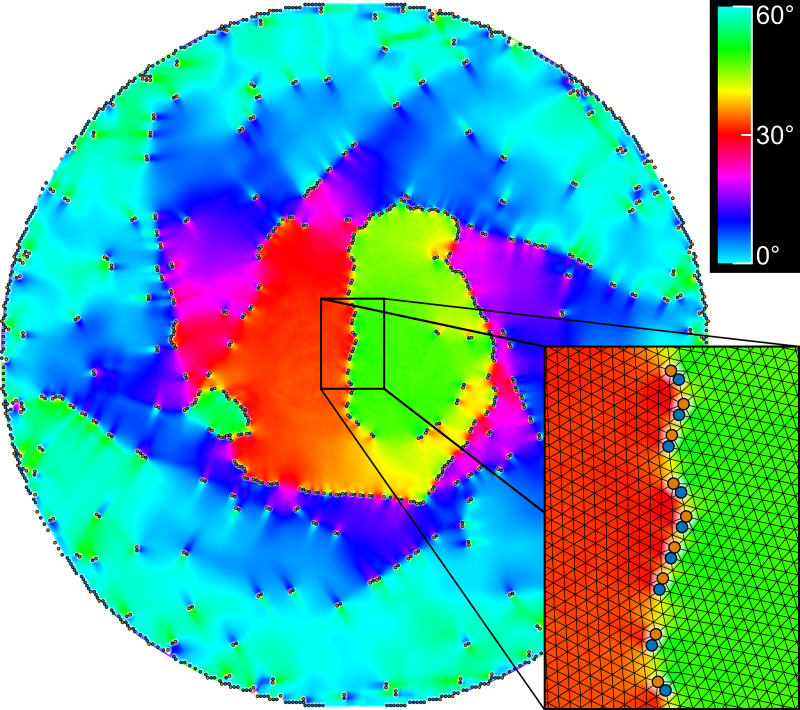}
		\caption{
			Overview of a full simulation of $\sim 35,000$ skyrmions. The magnification in the inset shows the skyrmion lattice and the structure of a domain wall. The two domains in the center of the sample (red/green) rotate relative to the boundary. Skyrmions on the boundary are pinned in order to realize an unperturbed lattice at large distances. Intermediate regions (blue/pink) dynamically deform to allow the relative rotation.
			The color gives the local orientation of the skyrmion lattice.
		}
		\label{fig:bigParticleSimZoom2}
	\end{figure}

	The skyrmion-skyrmion interaction $\vec F_{ij}=F(\vec R_i - \vec R_j)$ is calculated by using a fit to skyrmion-skyrmion potential which we obtained from micromagnetic simulations. We use $\vec F=- \vec \nabla V$ with $V(\vec r)\approx 2.31 e^{-(r/3.33)^{1.44}}$ within our unit system and set $m=1$, equilibrium skyrmion distance $\Delta_0\approx7.9$, $r_0\approx 266$, $v_0=6\cdot10^{-3}$ for the simulation shown in the Supplementary Movie. The value for $\mathcal{D}/G$ was extracted from the micromagnetic simulation of a single skyrmion, $\Delta_0$ from the simulation of a skyrmion lattice and $V(\vec{r})$ is a fit to a simulation of two skyrmions. All of these simulations were performed at $B=0.65$ in our units.
	Fig. \ref{fig:bigParticleSimZoom2} shows the result of such an analysis.
	
	
	We initialize our system on a perfect, defect-free hexagonal skyrmion lattice and fix the position of the particles at the circumference of radius $100 \Delta_0$, much larger than the typical diameter of the rotating domains in the center of the simulation. This disorder-free, high-symmetry situation allows to investigate the dynamics of domain formation in an idealized setup where pinning fixes the skyrmion lattice in the far distance while circular heat currents induce a rotation in the central part. As in the analysis of the experiment we perform Delaunay triangulation to identify lattice defects and calculate lattice orientation.
	Fig. \ref{fig:bigParticleSimZoom2} shows the result of such an analysis.
	
	Our simulations nicely reproduce that, under the shear stress arising from the current pattern, domains characterized by different rotational angle form. As in the experiment, they are separated by sharp domain walls characterized by the accumulation of 5-7 defects on the domain boundary, see Fig. \ref{fig:bigParticleSimZoom2}. Even for our idealized high-symmetry setup, the domain dynamics is a highly dynamic process characterized by continuous creation and destruction of domains. Whenever neighboring domains rotate with different angular velocities, the density of defects at the domain walls has to change as discussed in the main text. 
	Often far from the domain wall pairs of 5-7 defects are created and move with high speed relaxing the overall strain in the driven system. We believe that this motion is often too fast to be resolved in our experiments but in our simulation the motion of individual 5-7 defects is clearly a dominant mechanism for the dynamics of the whole system.
	The  defects tend to align into domain walls, as discussed. The influx of such defects can increase the defect density or reduce it when defects with opposite Burgers vectors annihilate. 
	The supplementary movie gives an overview of the dynamical, highly non-linear processes which accompany the rotation of the domains. Fig. 4 of the main text shows examplarily how defect motion and defect annihilation determines the domain wall dynamics.
	
	Our experimental observations and simulations have to  be distinguished from alternative scenarios for domain rotation. For example, one can envision a situation where crystalline domains rotating with different velocities are separated by a liquid region. For such a case domains can be stable and domain walls can be stationary. No large-distance defect motion or domain wall rearrangement is required. In contrast we find that the dynamics of the system is dominated by just such processes.

\end{document}